%% file: main.tex
\documentclass[10pt,leqno]{amsart}
\usepackage{graphicx}
\usepackage{indentfirst,csquotes}

\usepackage{amssymb,amsthm,amsmath}
\usepackage{xcolor,paralist,hyperref,titlesec,fancyhdr,etoolbox}

\titleformat{\section}{\normalfont\Large\bfseries\centering}{}{0pt}{}

\titleformat{\subsection}{\normalfont\large\bfseries}{}{0pt}{}

\hypersetup{ colorlinks=true, linkcolor=black, filecolor=black, urlcolor=black }

\usepackage{amsfonts}
\usepackage{subcaption}
\usepackage{algorithmic}
\usepackage{graphicx,wrapfig}
\usepackage{tabularx}
\usepackage{siunitx}
\usepackage{textcomp}
\usepackage{array} 
\usepackage{float}
\usepackage{adjustbox}
\usepackage{url}\usepackage{lscape}
 \usepackage{xspace}
\usepackage{makecell}
\usepackage{listings}
\usepackage{balance}
\usepackage{mdframed}
\usepackage{enumitem}
\usepackage{multirow}
\usepackage{rotating}
\usepackage{booktabs}
\usepackage{tikz}
\usepackage{pgfplots, pgfplotstable}
\pgfplotsset{
   compat=1.11,
   /pgf/number format/.cd,use comma,
   1000 sep = {\,},
       min exponent for 1000 sep = 4}
\usepackage{txfonts} 
\usetikzlibrary{positioning, shapes.geometric, arrows.meta}
\usepgfplotslibrary{groupplots}
\usetikzlibrary{patterns}

\lstdefinestyle{mystyle}{
    backgroundcolor=\color{white},   
    commentstyle=\color{green},      
    keywordstyle=\color{blue},       
    numberstyle=\tiny\color{gray},   
    stringstyle=\color{red},         
    basicstyle=\ttfamily\scriptsize,       
    breakatwhitespace=false,         
    breaklines=true,                 
    captionpos=b,                    
    keepspaces=true,                 
    numbers=left,                    
    numbersep=5pt,                   
    showspaces=false,                
    showstringspaces=false,          
    showtabs=false                   
}

\pgfplotsset{
   compat=1.11,
   /pgf/number format/.cd,use comma,
   1000 sep = {\,},
       min exponent for 1000 sep = 4}
\usepackage{txfonts}

\definecolor{mylightgray}{RGB}{224,224,224}

\newboolean{showcomments}
\setboolean{showcomments}{true}
\ifthenelse{\boolean{showcomments}}

\setboolean{showcomments}{true}
\ifthenelse{\boolean{showcomments}}
 { }

\newcommand{\finding}[1]{
	\setlength{\fboxrule}{1pt}
	\begin{center}\
		\noindent\fcolorbox{black}{gray!10}{
		\begin{minipage}{.92\linewidth}
			#1
		\end{minipage}
	}
	\end{center}
	\smallskip
}

\usepackage{pifont}
\newcommand{\wcircle}[1]{\ding{\numexpr171 + #1}}

\input{yaml}

\makeatletter
\renewcommand\section{\@startsection{section}{1}{\z@}%
  {-3.5ex plus -1ex minus -.2ex}
  {2.3ex plus .2ex}
  {\normalfont\Large\bfseries\centering}} 

\renewcommand\subsection{\@startsection{subsection}{2}{\z@}%
  {-2.5ex plus -1ex minus -.2ex}%
  {1.5ex plus .2ex}%
  {\normalfont\large\bfseries}} 
\makeatother

\lstdefinelanguage{puppet}{
  morekeywords={
    user, ensure, present, password, shell, managehome, file, owner, group, mode, directory
  },
  sensitive=false,
  morecomment=[l]{\#},  
  morestring=[b]'
}

\begin{document}
\title{Detection of Security Smells in IaC Scripts through Semantics-Aware Code and Language Processing} 

\author[]{
Aicha WAR
\and
Adnan A. RAWASS
\and
Abdoul K. Kabore
\and
Jordan Samhi
\and
Jacques KLEIN
\and
Tegawendé F. BISSYANDE \\
University of Luxembourg \\
}
\maketitle

\input{abstract}


\input{introduction}

\input{background}

\input{method}

\input{results}
\input{threats}
\input{related}
\input{conclusion}

\section{Data Availability}
In the interest of transparency and reproducibility, we provide the complete set of artifacts necessary to replicate and validate our findings. These include source code for preprocessing, model finetuning, and evaluation, final datasets of Ansible and Puppet code snippets, fine-tuned checkpoints for CodeBERT and Longformer, and all experimental configurations and logs with LLMs and previous work.  

All resources are organized in the following anonymous GitHub repository: 
\url{https://anonymous.4open.science/r/semantics-aware-iac-analysis-0F76}.

\balance
\bibliographystyle{IEEEtran} 
\bibliography{references.bib}
\end{document}

%% file: abstract.tex
\begin{abstract}
Infrastructure as Code (IaC) automates the provisioning and management of IT infrastructure through scripts and tools, streamlining software deployment. Prior studies have shown that IaC scripts often contain recurring security misconfigurations, and several detection and mitigation approaches have been proposed. Most of these rely on static analysis, using statistical code representations or Machine Learning (ML) classifiers to distinguish insecure configurations from safe code.

In this work, we introduce a novel approach that enhances static analysis with semantic understanding by jointly leveraging natural language and code representations. Our method builds on two complementary ML models: CodeBERT, to capture semantics across code and text, and LongFormer, to represent long IaC scripts without losing contextual information. We evaluate our approach on misconfiguration datasets from two widely used IaC tools, Ansible and Puppet. To validate its effectiveness, we conduct two ablation studies (removing code text from the natural language input, and truncating scripts to reduce context) and compare against four large language models (LLMs) and prior work. Results show that semantic enrichment substantially improves detection, raising precision and recall from 0.46 and 0.79 to 0.92 and 0.88 on Ansible, and from 0.55 and 0.97 to 0.87 and 0.75 on Puppet, respectively.
\end{abstract}

%% file: introduction.tex
\section{Introduction}
\label{sec:introduction}

Infrastructure as Code (IaC) has become a cornerstone of modern IT operations, facilitating the automated and remote provisioning and management of data center resources through scripts, thereby significantly reducing dependence on manual interventions. This paradigm shift encompasses the orchestration of diverse tools and configuration files to support a wide array of physical hardware, cloud resources, and virtual machines \cite{wittig2016aws}. Popular IaC frameworks, which leverage tools such as Puppet and Ansible, are instrumental in defining and deploying infrastructure, extensively interacting with servers and cloud environments. 
The widespread adoption of IaC is driven by its ability to enhance agility, consistency, and scalability in infrastructure management, making it an indispensable practice in contemporary software development and operations.

While IaC scripts undeniably boost efficiency, their inherent power also introduces a critical vulnerability: even minor mistakes can precipitate severe security risks. 
Misconfigurations, such as the inadvertent use of HTTP instead of HTTPS, reliance on deprecated cryptographic algorithms like SHA-1, or the deployment of unsupported operating systems, can lead to significant security breaches and substantial financial losses. 
Illustrative examples include Amazon's reported loss of \$150 million due to security vulnerabilities within its S3 service \cite{saavedra2022glitch}, Capital One's data breach resulting from improperly configured AWS S3 buckets \cite{volkov2022capitalone}, and Uber's exposure of unencrypted secret keys in a public repository on GitHub in 2017 (Uber, 2017). These incidents underscore the urgent and persistent need for robust, accurate, and automated detection mechanisms to identify and mitigate security misconfigurations in IaC scripts before they can be exploited.

The development of effective tools for accurate and automated detection of security misconfigurations in IaC scripts is paramount. Existing approaches to detect misconfigurations often rely on static analysis~\cite{chiari2022static,konala2023sok, chess2004static}, rule-based systems~\cite{saavedra2022glitch,hassan2022code}, or traditional machine learning techniques~\cite{borovits2020deepiac,borovits2022findici}. 
Although these methods have shown some success, they frequently struggle with the nuanced nature of IaC, which blends declarative syntax with operational intent. 
Traditional detection methods may suffer from high false positive rates due to a lack of semantic understanding~\cite{chiari2022static,schwarz2018code}, or they may fail to identify novel or context-dependent misconfigurations that deviate from predefined rules~\cite{rahman2018characterizing,dalla2021within}. 
Furthermore, many current tools lack the adaptability to keep up with the rapid evolution of IaC frameworks and cloud services, leaving gaps in their detection capabilities. 
This limitation highlights a critical need for more sophisticated approaches that can interpret the intended meaning and contextual relationships within IaC code, moving beyond simple syntactic checks to identify subtle yet critical security flaws.

This study addresses the aforementioned challenges by significantly enhancing misconfiguration detection through the innovative integration of advanced Machine Learning (ML) approaches with state-of-the-art large language models (LLMs) such as GPT-3.5, GPT-4, CodeBERT, Longformer, Starcoder-2, and LLaMA-3. Our approach specifically explores how combining natural language processing (NLP) techniques with code processing can yield a deeper, more semantic understanding of IaC scripts. This focus on semantic analysis allows us to interpret the intended meaning and contextual relationships within IaC code, which is crucial to distinguish between safe IaC code and actual security misconfigured ones. To achieve this, we first collected extensive and high-quality datasets of Ansible and Puppet (two popular IaC tools) misconfigured IaC code snippets derived from the literature~\cite{dalla2021within,rahman2018characterizing}. Our Ansible dataset contains IaC code snippets with text (code comments, task descriptions), while our Puppet dataset only contains long IaC code snippets with full classes for more analysis context. The novelty in this paper is that we consider both code and text as forms of natural language in our semantic-aware analysis of IaC scripts. Second, we fine-tuned CodeBERT and Longformer models by retraining them on our collected datasets. This fine-tuning process was designed to significantly enhance the models' ability to effectively distinguish between correct IaC code and security misconfigurations by representing the features of our collected code snippets and then by classifying the security-misconfigured ones from correct code snippets.

Second, we performed two ablation experiments to evaluate the impact of semantic-aware analysis on the accurate detection of security misconfigurations in IaC scripts. Specifically, we created two new datasets by removing all text from our Ansible dataset of code snippets and we reduce the length of our Puppet code snippets for less natural language context. Then we tested our fine-tuned models on our ablated datasets and reported the results. Finally, we carried out a comparison experiment with prior work and LLMs. Our goal was to test our fine-tuned models on prior work datasets and leverage LLMs (also efficient in NLP and code processing) for static analysis on our collected datasets to evaluate and validate our work in comparison with existing and modern techniques.


The main contributions of our study are as follows:
\begin{itemize}
    \item We construct high-quality datasets of Ansible and Puppet code snippets, including both security-misconfigured and correctly implemented IaC code, along with ablated variants for controlled evaluation.  
    \item We develop an ML-based approach leveraging CodeBERT and LongFormer that captures the semantics of IaC scripts and outperforms prior work in detecting IaC security misconfigurations.  
    \item We provide a comparative evaluation against popular LLMs for static analysis of IaC scripts, demonstrating that our semantic-aware models consistently achieve superior performance over modern general-purpose solutions.  
\end{itemize}

%% file: background.tex
\section{Background}
\label{sec:background}

In this section, we present the foundations of Infrastructure as Code.  
We review existing analysis techniques and discuss the role of modern machine learning models (including large language models).  
This contextualization provides the necessary background to better understand our approach and contributions.  

\subsection{Infrastructure as Code}

Infrastructure as Code is a cornerstone of modern DevOps practice that enables the provisioning and configuration of infrastructure using machine-readable script files.
Tools such as Ansible, Puppet, Terraform, and SaltStack empower practitioners to define and manage infrastructure resources, such as virtual machines, networks, and services, entirely in code rather than through manual operations. 
However, it also introduces new challenges and risks. 
As IaC scripts evolve, they may suffer from defects, configuration drift, or misconfigurations, potentially impacting system reliability, security, and performance. 
Moreover, their declarative and domain-specific nature makes their analysis distinct from traditional software programs.

\begin{figure}[h]
\begin{lstlisting}[language=ruby, basicstyle=\tiny, numbers=left, xleftmargin=3em, caption={Example Puppet script with plaintext password and overly permissive directory permissions.}, captionpos=b, label={fig:background_example}]
# Creates a user 'deploy' with a weak password hash
user { 'deploy':
  ensure     => present,
  password   => 'password',  # Security smell: plaintext or weak password
  shell      => '/bin/bash',
  managehome => true,
}

# Grants overly permissive access to a sensitive directory
file { '/opt/deploy':
  ensure  => 'directory',
  owner   => 'deploy',
  group   => 'deploy',
  mode    => '0777',  # Misconfiguration: excessive directory permissions
}
\end{lstlisting}
\end{figure}

Listing \ref{fig:background_example} illustrates a Puppet script snippet containing two security misconfigurations: 
\wcircle{1} the use of a plaintext password and 
\wcircle{2} the assignment of overly permissive directory permissions. 
These examples highlight the type of vulnerabilities that IaC scripts can introduce when improperly configured.
In this work, we aim to improve the accuracy of static analysis techniques for detecting such misconfigurations in IaC scripts.
We focus, in particular, on Ansible and Puppet as representative use cases.

\subsection{Static Analysis of IaC Scripts}

To address quality and reliability concerns in IaC scripts, several efforts have introduced static analysis~\cite{chess2004static} techniques for detecting misconfigurations and anti-patterns~\cite{rahman2018characterizing,dalla2021within}. 
Early research focused on mining product and process metrics from version control systems, including script length, change frequency, and author activity, to build defect prediction models~\cite{dalla2021within}. 
These statistical approaches~\cite{rahman2018characterizing} have proven valuable, as demonstrated by studies correlating such metrics with fault-proneness in IaC scripts. The effectiveness of static analysis also strongly depends on how code is represented for machine processing. Code representation refers to the transformation of raw scripts into abstract formats, such as trees, graphs, or vectors that can be analyzed by algorithms or machine learning models. For IaC scripts, choosing an appropriate representation is challenging due to their declarative structure and configuration-centric semantics. Traditional approaches rely on handcrafted features or abstract syntax trees (ASTs) that encode syntactic relationships~\cite{begoug2024terrametrics,borovits2020deepiac,schwarz2018code,konala2023sok,saavedra2022glitch}. More advanced representations include control/data-flow graphs and dependency graphs, which attempt to capture resource relationships and execution order~\cite{opdebeeck2023control,bykov2025infrastructure,jeong2021dataflow,bacon2014information}. Recently, learning-based approaches have introduced neural code embeddings that map code tokens or structures into dense vectors, enabling models to capture latent patterns across scripts~\cite{borovits2022findici,diaz2024towards,chiari2022static}. These embeddings can support tasks such as defect prediction, similarity detection, and script classification, and are more adaptable to the complexities of IaC.

\subsection{Natural Language and Code Processing Models}

Pre-trained language models from the NLP domain have shown strong potential in software engineering tasks involving both natural language and code. Models such as \textit{CodeBERT}, trained jointly on code and textual descriptions, can capture contextual and semantic relations within scripts, even in the presence of minimal documentation. Additionally, transformer-based architectures like \textit{LongFormer}, designed for long sequences, are especially suitable for analyzing lengthy IaC scripts that exceed standard model input limits. These models are widely studied in the literature~\cite{borovits2020deepiac,saavedra2022glitch,gunawat2025ai,goyal2024detecting} and provide a unified representation of code structure, variable usage, and configuration semantics. Furthermore, recent studies have discovered that code is also a natural language~\cite{10.5555/2337223.2337322}. Hence, providing more code context can be assimilated to providing more semantic context as a code comment would. In our work, we build on these advances to improve semantic understanding, code context, and misconfiguration detection in IaC scripts, extending beyond prior metric-based approaches.

\subsection{Large Language Models for Misconfiguration Detection}

Large Language Models (LLMs) have recently shown good capabilities in analyzing source code and configuration files. Trained on large corpora of code and natural language, these models can capture complex syntactic and semantic patterns, making them capable of identifying misconfigurations in IaC scripts~\cite{wen2025llm, malul2024genkubesec, minna2025analyzing}. Unlike traditional static analysis tools, LLMs can leverage both code and accompanying natural language (e.g., comments, variable names) to infer configuration intent. This allows them to detect subtle issues such as insecure defaults, weak permissions, or missing encryption settings. Recent studies show that LLMs can highlight security smells, suggest remediations, and even generate corrected configurations, often with accuracy comparable to or better than static tools~\cite{li2024enhancing, li2023assisting}. While LLMs offer flexibility and contextual reasoning, they also pose challenges, including potential hallucinations and lack of formal guarantees. Nonetheless, their ability to generalize across diverse IaC patterns makes them a valuable complement to rule-based detection. In this paper, we establish a comparative experiment between our fine-tuned CodeBert and LongFormer models and four LLMs to assess their capabilities for misconfiguration detection in IaC scripts and evaluate our work.


%% file: method.tex
\section{Methodology}
\label{sec:methodology}

In this work, we propose a methodology to enhance the detection of security misconfigurations in IaC scripts.  
Our approach builds on recent advances in machine learning and large language models, aiming to capture the semantics of IaC beyond purely syntactic checks.  
By combining natural language elements (e.g., task descriptions and comments) with code analysis, we seek to provide a deeper understanding of IaC scripts and thereby improve the robustness and accuracy of misconfiguration detection.  

Our methodology is designed to address the following research questions (RQs):

\begin{description}
    \item[RQ1:] \textbf{How do the integration of natural language processing with code analysis techniques affect the detection of security misconfigurations in IaC scripts?}  
    This question investigates first, whether combining text elements (e.g., task descriptions, comments) with code representations leads to a more accurate identification of security misconfigurations than analyzing code alone. Second, it investigates how more code context (longer code sequences) can influence the detection of these IaC misconfigurations, code being considered as a form of natural language.  

    \item[RQ2:] \textbf{To what extent does improved semantic analysis of IaC scripts enhance the detection of security misconfigurations?}  
    Here, we examine the role of capturing deeper semantic relationships within IaC scripts, evaluating whether richer contextual understanding contributes to more precise detection.  

    \item[RQ3:] \textbf{How does our approach perform compared to LLMs?}  
    This question benchmarks our method against four popular LLM versions to assess its effectiveness in detecting security misconfigurations in IaC scripts.  
\end{description}

In the following section, we detail the experimental setup for addressing each RQ.

\section{Experimental Setup}
\label{sec:experimental_setup}

To answer our RQs, we designed a series of controlled experiments and compared the results with prior work.
This section presents the setup we used for each RQ.

\subsection{Setup for RQ1: NLP Integration Impact}
\label{sec:rq1}

\noindent
\textbf{Data Collection.}  
To evaluate the impact of integrating natural language processing with code analysis, we relied on two complementary datasets: 
\wcircle{1} Ansible (rich in natural language content) and 
\wcircle{2} Puppet (largely code-only).  
This design allows us to compare the contribution of natural language context, such as task descriptions and comments, to the detection of security misconfigurations.  

For Ansible, we collected misconfigured code snippets from the validated dataset of Dalla Palma et al.~\cite{dalla2021within}.
In total, we extracted \num{2759} misconfiguration-fix commit pairs and reconstructed both the before and after versions, enabling the models to learn differences between faulty and corrected code.  
The final dataset consists of \num{3066} snippets, split into \num{2146} for fine-tuning, \num{613} for validation, and \num{307} for testing.  
An example is shown in Listing~\ref{fig:ansible_example}, where the improper use of the ``shell'' module and a hardcoded file mode (``0750'') introduce security and portability risks.  
A distinctive feature of this dataset is the presence of natural language text (e.g., task names, descriptions, and comments), which we hypothesize provides valuable semantic context for distinguishing misconfigurations.  

\begin{figure}[h]
    \begin{lstlisting}[language=yaml, basicstyle=\tiny, numbers=left, xleftmargin=3em, caption={Ansible code snippet configuring Nextcloud, showing misconfigurations through hardcoded mode and unsafe shell use.}, captionpos=b, label={fig:ansible_example}]
- name: "[NC] - Set Nextcloud settings in config.php"
  shell: php occ config:system:set {{ item.name }} --value="{{ item.value }}"
  with_items:
    - "{{ nextcloud_config_settings }}"

- name: "[NC] - Set Redis Server"
  command: php occ config:system:set {{ item.name }} --value="{{ item.value }}"
  with_items:
    - "{{ nextcloud_redis_settings }}"
  when: nextcloud_install_redis_server == True
  mode: 0750
    \end{lstlisting}
\end{figure}

For Puppet, we built our dataset from the validated dataset of Rahman et al.~\cite{rahman2018characterizing}.
The dataset contains security-misconfigured code snippets labelled as 1 and clean code snippets labelled as 0.
The final dataset contains \num{1958} snippets: \num{1370} for fine-tuning, \num{392} for validation, and \num{196} for testing.  
Unlike Ansible, Puppet scripts are typically long code sequences with minimal natural language content, making them particularly suitable for models designed to handle extended contexts such as LongFormer.  
Listing~\ref{fig:puppet_example} illustrates a Puppet manifest that embeds sensitive values (e.g., passwords and connection strings) directly in the configuration, creating severe risks of credential leakage.

\begin{figure}[h]
  \begin{lstlisting}[language=ruby, xleftmargin=2em, basicstyle=\tiny, numbers=left, caption={Puppet Code Snippet Deploying OpenStack Trove Components with Insecure Secret Handling}, captionpos=b, label={fig:puppet_example}]
class { '::trove::client': }

class { '::trove::keystone::auth':
  admin_address   => '10.0.0.1',
  internal_address => '10.0.0.1',
  public_address  => '10.0.0.1',
  password        => 'verysecrete',
  region          => 'openstack',
}

class { '::trove::db::mysql':
  password       => 'dbpass',
  host           => '10.0.0.1',
  allowed_hosts  => '10.0.0.1',
}

class { '::trove':
  database_connection     => 'mysql://trove:secrete@10.0.0.1/trove?charset=utf8',
  rabbit_hosts            => '10.0.0.1',
  rabbit_password         => 'secrete',
  nova_proxy_admin_pass   => 'novapass',
}

class { '::trove::api':
  bind_host         => '10.0.0.1',
  auth_url          => 'https://identity.openstack.org:5000/v2.0',
  keystone_password => 'verysecrete',
}

class { '::trove::conductor':
  auth_url => 'https://identity.openstack.org:5000/v2.0',
}

class { '::trove::taskmanager':
  auth_url => 'https://identity.openstack.org:5000/v2.0',
}
  \end{lstlisting}
\end{figure}

\noindent
\textbf{Data Preprocessing.}
Before fine-tuning, we prepared and standardized the datasets to maximize learning effectiveness~\cite{jurafsky2008}.  
The following steps were applied:

\begin{itemize}
    \item \textit{Lowercasing:} Converted all characters to lowercase for consistency~\cite{siavvas2021}.
    \item \textit{Special Character Filtering:} Removed non-semantic characters to reduce noise~\cite{harzevili2023}: some punctuations (such as .,!) and replaced any characters that cannot be converted to UTF-8 with empty strings.
    \item \textit{Single-Line Conversion:} Flattened code into single-line sequences while preserving syntactic correctness, improving tokenization efficiency.
\end{itemize}

\noindent
\textbf{Model Fine-Tuning.}
We fine-tuned CodeBERT and LongFormer to classify misconfigured versus corrected IaC snippets.  
CodeBERT was primarily used for Ansible scripts, while both CodeBERT and LongFormer were applied to Puppet.  
Fine-tuning was conducted using \num{1}–\num{4} NVIDIA A100 GPUs (\num{128} GB), with batch sizes adapted to sequence lengths: \num{64} and \num{128} for CodeBERT, and \num{8} and \num{16} for LongFormer.  
To prevent overfitting and ensure stable convergence, we employed \num{8}-fold cross-validation, AdamW optimization~\cite{kingma2015adam}, linear learning rate warm-up (first \num{10}\% of steps), weight decay (\num{0.01} for CodeBERT and LongFormer; \num{0.001} for fine-tuned LongFormer), and early stopping if validation loss stagnated for five consecutive epochs.  
We used RoBERTa-based classification for CodeBERT and LongformerForSequenceClassification for LongFormer.  

\noindent
\textbf{Validation.}
Using our test datasets, we measured the ability of our fine-tuned models to distinguish between misconfigured (label ``\num{1}'') and corrected (label ``\num{0}'') code snippets.  
The inclusion of both faulty and fixed versions in our datasets allowed the models to learn the semantic differences underlying security fixes.  

\noindent
\textbf{Evaluation Metrics.}

To evaluate performance, we relied on Precision, Recall, and F1-score, given their effectiveness in imbalanced classification scenarios.  
Precision measured the correctness of positive predictions, while Recall assessed the model’s ability to capture actual misconfigurations.  
The F1-score provided a balanced measure of both dimensions.  

\noindent
\textbf{Baselines from the Literature.}
Prior studies on IaC misconfiguration detection rely on statistical feature extraction techniques such as Term Frequency–Inverse Document Frequency (TF-IDF) and Bag of Words (BoW)~\cite{rahman2018characterizing,dalla2021within}, combined with Random Forest classifiers.  
These models convert IaC scripts into numerical representations but fail to capture semantic relationships or contextual intent in the code.  
For a fair comparison, we reproduced their setups on our datasets:
\begin{itemize}[noitemsep,topsep=0pt,left=0pt]
    \item For Ansible, we directly reused the dataset from Dalla Palma et al.~\cite{dalla2021within}, ensuring that TF-IDF, BoW, and Random Forest were applied under the same conditions as in the original work.  
    \item For Puppet, we reconstructed the approach of Rahman et al.~\cite{rahman2018characterizing}, who originally analyzed three separate datasets (OpenStack, Mozilla, Wikimedia).  
Since our Puppet dataset merges these sources, we replicated their design on the unified dataset while applying the same train/validation/test splits to avoid overfitting~\cite{howard2018ulmfit}.  
\end{itemize}

\subsection{Setup for RQ2: Semantic Analysis Enhancement}

\noindent
\textbf{Data Collection and Ablation Design.}  
To evaluate the impact of richer semantic context on detecting misconfigurations, we performed an ablation experiment on both Ansible and Puppet datasets.  
For Ansible, we created a reduced dataset by removing all natural language elements (e.g., task descriptions and comments), leaving only executable code.  
For Puppet, we shortened each snippet to preserve only the misconfigured instruction(s), along with \num{3} lines of preceding context and \num{2} lines of subsequent context.  
This design allowed us to test whether limiting contextual information degrades model performance compared to the full datasets.  

We automated the cleaning process using GPT-3.5-Turbo to ensure systematic removal and reduction of code content.  
To guarantee data quality, all processed snippets were then manually validated, and no output was discarded. The LLM reliably provided Ansible code without text and maintained the code logic. 
Prompt 1 and Prompt 2 show the prompts provided to the LLM for cleaning the Ansible and Puppet datasets, respectively.

\begin{mdframed}[style=niceframe, linecolor=black, frametitle={Prompt 1},frametitlealignment=\centering, userdefinedwidth=0.95\columnwidth, align=center, font=\scriptsize]
\vspace{-.4cm}
\textit {``You are an expert in Ansible YAML task cleanup.
Your job is to remove ONLY the parameters that do not contain 
executable logic, specifically those with natural language text. 
Additionally, remove all comments. Return ONLY the cleaned YAML 
snippet. Do not add any explanation.
Here is the YAML snippet:
[CODE\_SNIPPET]''}
\label{fig:ansible_cleaning}
\end{mdframed}

\begin{mdframed}[style=niceframe, linecolor=black, frametitle={Prompt 2},frametitlealignment=\centering, userdefinedwidth=0.95\columnwidth, align=center, font=\scriptsize]
\vspace{-.4cm}
\textit {``You are given a Puppet code snippet. Your task is to extract a 
reduced version of the script containing only the security
misconfigured instruction(s) along with a small code context 
around them. Specifically, include the misconfigured line(s) as 
well as 3 lines before and 2 lines after to retain minimal
readability. Ensure the extracted script maintains correct
indentation and valid Puppet syntax. 
Return only the reduced code block.
\\
Example Input: \\
class \{ '::trove': \\
\indent database\_connection =$>$ 'mysql://trove:secrete@10.0.0.1,\\
\indent rabbit\_hosts            =$>$ `10.0.0.1', \\
\indent rabbit\_password         =$>$ `secrete', \\
\indent nova\_proxy\_admin\_pass   =$>$ `novapass', \\
\}
\\
Expected Output: \\
class \{ '::trove': \\ 
\indent database\_connection     =$>$ `mysql://trove:secrete@10.0.0.1, \\
\indent rabbit\_hosts            =$>$ `10.0.0.1',\\
\indent rabbit\_password         =$>$ `secrete',\\
\}
\\
Here is the Puppet code snippet:
[CODE\_SNIPPET]''}
\label{fig:puppet_cleaning}
\end{mdframed}

\medskip

\noindent
\textbf{Data Preprocessing.}  
The ablated datasets underwent the same preprocessing steps described in Section~\ref{sec:rq1}, including lowercasing, filtering of non-semantic characters, and single-line conversion for tokenization efficiency.  
As with the original datasets, CodeBERT was applied to Ansible, while LongFormer handled Puppet code sequences.  

\noindent
\textbf{Models.}  
For Ansible, we evaluated our already fine-tuned CodeBERT model on the reduced dataset.  
For Puppet, we tested our fine-tuned LongFormer models on the shortened snippets.  
The same fine-tuning infrastructure and hyperparameter settings as in RQ1 were used to ensure comparability.  

\noindent
\textbf{Evaluation Metrics.}
\label{sec:metrics}
We measured Precision, Recall, and F1-score to assess model effectiveness.  
By comparing the results of the ablated datasets against those from the original datasets, we evaluated the degree to which semantic context improves the detection of IaC misconfigurations.  

\subsection{Setup for RQ3: Comparison with LLMs}

To assess the performance of our method, we conducted a comparative study involving our fine-tuned models (CodeBERT and LongFormer) and LLMs widely used in recent work. We selected four LLMs for benchmarking: GPT-4-Turbo, GPT-3.5-Turbo, Starcoder-2, and LLaMA-3.2.  
These models were chosen for their ability to process both code and natural language, and because they were trained on corpora that include languages relevant to IaC: Python (for YAML-like Ansible scripts) and Ruby (for Puppet).
\begin{itemize}[noitemsep,topsep=0pt,left=0pt]
    \item GPT-4-Turbo and GPT-3.5-Turbo each contain \num{175} billion parameters and a \num{4096}-token context window, offering advanced reasoning abilities.  
    \item Starcoder-2, with \num{7} billion parameters, is open-source and trained on millions of YAML and Ruby files, providing strong domain coverage.   
    \item LLaMA-3.2, with \num{8} billion parameters and an \num{8000}-token context window, offers a lightweight open-source alternative.  
\end{itemize}

Each LLM was evaluated using carefully designed prompts to detect security misconfigurations by comparing a vulnerable code snippet (label ``\num{1}'') with its corrected version (label ``\num{0}'').  
The prompts required the models to identify misconfigurations and output associated CWE identifiers, as illustrated below for Ansible and Puppet respectively:

\begin{mdframed}[style=niceframe, linecolor=black, frametitle={Prompt 3},frametitlealignment=\centering, userdefinedwidth=0.95\columnwidth, align=center, font=\scriptsize]
\vspace{-.4cm}
\textit {`You are a security expert in Ansible scripts. Analyze the code 
below and report only security misconfigurations with real 
CWE IDs. Make assumptions based on the following inputs:
Vulnerable code: [VULNERABLE\_CODE]
Fixed code: [FIXED\_CODE]''}
\end{mdframed}

\begin{mdframed}[style=niceframe, linecolor=black, frametitle={Prompt 4},frametitlealignment=\centering, userdefinedwidth=0.95\columnwidth, align=center, font=\scriptsize]
\vspace{-.4cm}
\textit {``You are a security expert in Puppet scripts. Analyze the code 
below and report only security misconfigurations with real 
CWE IDs. Make assumptions based on the following inputs:
Vulnerable code: [VULNERABLE\_CODE]
Fixed code: [FIXED\_CODE]''}
\end{mdframed}

To reduce hallucinations, we required CWE-based outputs, which allowed manual verification of misconfigurations and ensured alignment with recognized vulnerability categories.  

\noindent
\textbf{Our Models.}
We included our fine-tuned CodeBERT and LongFormer models in the comparison.  
They were trained on the full Ansible and Puppet datasets described in Section~\ref{sec:rq1}, using the preprocessing and fine-tuning pipeline detailed earlier.  
CodeBERT was applied to both Ansible and Puppet, while LongFormer was primarily evaluated on Puppet due to its ability to process long sequences without truncation.  

\noindent
\textbf{Evaluation Protocol.}
All approaches were tested on the same labeled datasets, with train/validation/test splits aligned across methods to ensure comparability.  
For the LLMs, predictions were collected via prompt-based inference, and outputs were manually validated when CWE labels were ambiguous.  
Traditional ML baselines and our models were evaluated using their respective classifiers (Random Forest for TF-IDF/BoW; RoBERTa for CodeBERT; LongformerForSequenceClassification for LongFormer).  

\noindent
\textbf{Metrics.}
We relied primarily on Precision, Recall, and F1-score to measure detection accuracy. 
These metrics allowed us to quantify the relative advantages of semantic-aware models (our method and LLMs).

%% file: results.tex
\section{Experimental Results}
\label{sec:results}

We present the results of our study in this section. First, we evaluate the effectiveness of enhancing semantic understanding in static analysis for identifying security misconfigurations in IaC scripts (RQ1~\ref{subsec:rq1}). Next, we investigate how semantic analysis influences the detection of these misconfigurations through ablation experiments (RQ2~\ref{subsec:rq2}). Finally, we compare the performance of LLMs with that of our fine-tuned ML models in detecting security issues in IaC scripts (RQ3~\ref{subsec:rq3}).

\subsection{RQ1: NLP for Detecting IaC Misconfigurations}
\label{subsec:rq1}

We compare the performance of our fine-tuned models on the collected Puppet and Ansible datasets, using evaluation metrics such as precision, recall, and F1-score. To obtain a comprehensive assessment, we benchmarked our approach with traditional machine learning methods from the literature. The analysis highlights the strengths and weaknesses of each technique in identifying security misconfigurations in IaC scripts.

\subsubsection{LongFormer vs. TF-IDF and BoW on Puppet Dataset}
\label{subsec:Longformer_CodeBERT}

We evaluated LongFormer against traditional text-mining techniques (TF-IDF and BoW) on the Puppet dataset, which contains long code snippets. As shown in Table~\ref{tab:model_performance_puppet}, LongFormer achieved the best overall performance, with the highest precision (0.87) and F1-score (0.79), outperforming TF-IDF (0.70) and BoW (0.70).

\begin{table}[htbp]
\centering
\caption{Overall Performance of Previous Work and LongFormer on Puppet Dataset for the Detection of Security Misconfigurations}
\label{tab:model_performance_puppet}
\resizebox{\linewidth}{!}{
\begin{tabular}{lcccccc}
\hline
\textbf{Metric} & \textbf{TF-IDF} & \textbf{BoW} & \textbf{LF B16} & \textbf{LF B8} & \textbf{FT LF B8} & \textbf{FT LF B16} \\ \hline

\textbf{Precision} & 0.70 & 0.70 &
0.73 & 0.72 & 0.78 & 0.87 \\ \hline

\textbf{Recall} & 0.73 &0.74 &
0.68 & 0.72 & 0.77 & 0.75 \\ \hline

\textbf{F1-score} &0.72 &0.72 &
0.70 & 0.71 & 0.76 & 0.79 \\ \hline
\multicolumn{6}{l}{\small $^{\mathrm{*}}$ LF= Longformer, FT= Fine-Tuned, B8= Batch size 8, B16= Batch size 16.}
\end{tabular}
}
\end{table}

Although TF-IDF and BoW achieved slightly higher recall (0.73 and 0.74, respectively), their precision was lower than that of LongFormer, resulting in less balanced performance for misconfiguration detection.
These results highlight LongFormer’s effectiveness in analyzing Puppet scripts and its advantage in handling longer sequences \cite{jiang2024stagedvulbert}.
In the following section, we examine the impact of weight decay optimization on convergence speed and detection efficiency.

\subsubsection{Hyperparameter Tuning on LongFormer Performance (Batch Size 8) }
\label{par:hyperparam}

\begin{figure}[!htbp]
    \centering
    \includegraphics[width=\linewidth]{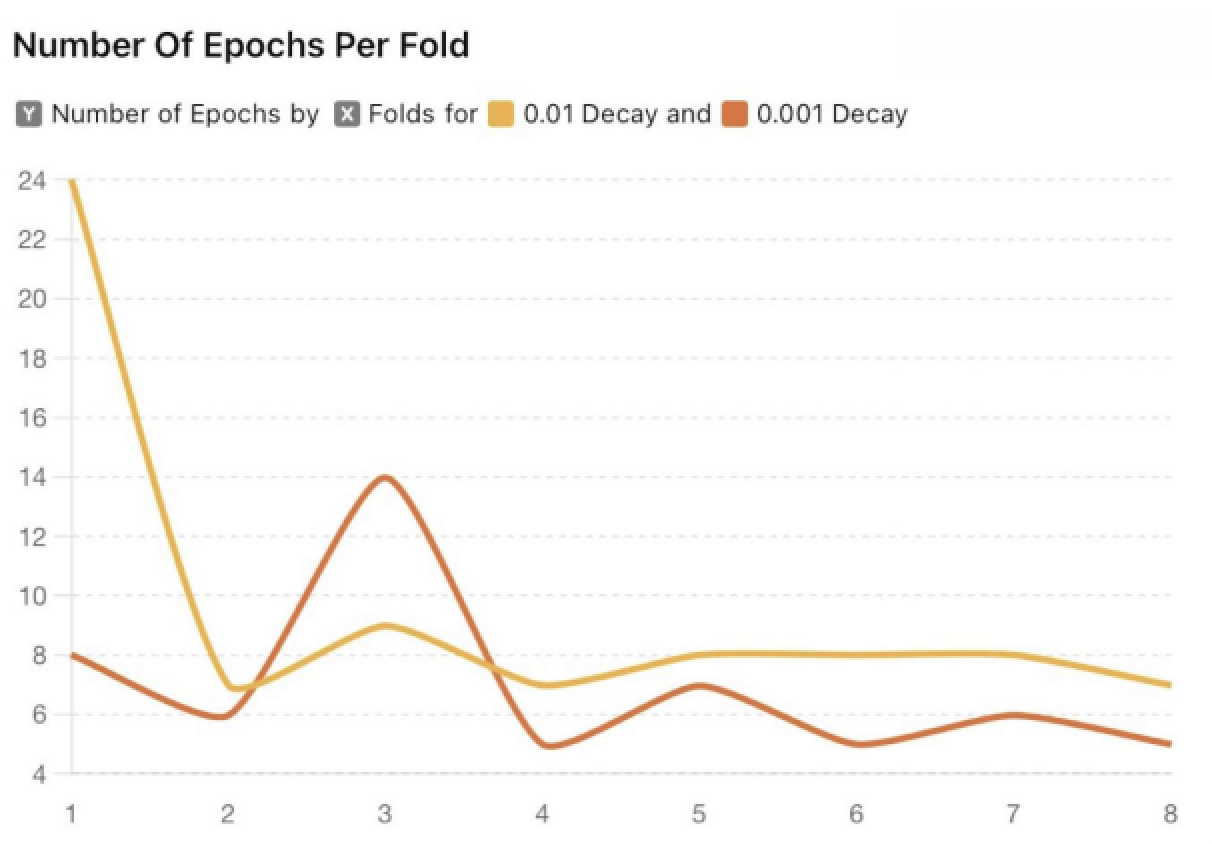}
    \caption{Epochs per Fold with Weight Decay in LongFormer Fine-tuning on Complete Puppet Dataset }
    \label{fig:Weight Decay}
\end{figure}

We examined the effect of weight decay during hyperparameter tuning.
A high decay rate of \num{0.01} accelerated convergence but resulted in unstable fine-tuning and limited learning performance.
In contrast, a lower decay rate of \num{0.001} reduced the number of epochs required for each fold to converge.
Notably, fine-tuning under lower decay values was considerably more stable, leading to improved learning outcomes.
These observations indicate that weight decay plays a critical role in balancing convergence speed and fine-tuning stability of pretrained models, thereby enhancing the effectiveness of detecting security misconfigurations in IaC, as illustrated in Figure~\ref{fig:Weight Decay}.

The median results of our experiments with different batch sizes using LongFormer are presented in Table~\ref{tab:model_performance_puppet}.
Through fine-tuning and hyperparameter optimization, we achieved notable improvements in performance metrics.
For example, the precision increased from \num{0.73} to \num{0.87} for a batch size of \num{16} and from \num{0.72} to \num{0.78} for a batch size of \num{8}.
Similarly, the recall improved from \num{0.72} to \num{0.77} for a batch size of \num{8}.
These results demonstrate that hyperparameter tuning significantly enhanced LongFormer’s effectiveness on Puppet scripts for detecting security-misconfigured code snippets.

\subsubsection{A Comparison of CodeBERT with Random Forest and BoW on the Ansible Dataset}
\label{subsec:CodeBERT_original}

We compare our fine-tuned CodeBERT model against a traditional Random Forest (RF) classifier trained with Bag of Words (BoW) features, as reported in prior work~\cite{dalla2021within}, using the same dataset of \num{3478} Ansible code snippets for the detection of security misconfigurations.

As shown in Table~\ref{tab:performance_comparison_ansible}, CodeBERT achieves substantial improvements across all evaluated metrics. Precision increased from 0.77 in the previous approach to 0.92, indicating a significant reduction in false positives. Recall also improved, from 0.84 to 0.88, demonstrating that CodeBERT captures a larger proportion of actual misconfigurations. Consequently, the F1-score increased from 0.77 to 0.90, highlighting a more balanced and robust overall performance.

\begin{table}[htbp]
\centering
\small
\caption{Performance Comparison between Random Forest+Bag of Words and CodeBERT on Ansible Dataset for the Detection of Security Misconfigurations}
\label{tab:performance_comparison_ansible}
\resizebox{\linewidth}{!}{
\begin{tabular}{lcc}
\hline
\textbf{Metric} & \textbf{RF + BoW} & \textbf{CodeBERT} \\ \hline
Precision & 0.77   & 0.92 \\
Recall    & 0.84   & 0.88 \\
F1-score  & 0.77   & 0.90 \\ \hline
\multicolumn{3}{l}{\small $^{\mathrm{*}}$  RF = Random Forest, BoW = Bag of Words, CB = CodeBERT.}
\end{tabular}
}
\end{table}

The most pronounced improvement is in precision, which suggests that CodeBERT’s ability to understand the contextual semantics of code helps it distinguish between benign and vulnerable configurations more effectively than traditional feature-based models. By leveraging pre-trained embeddings and fine-tuning on the Ansible dataset, CodeBERT generalizes better to unseen patterns, improving the detection of subtle security misconfigurations that BoW-based RF models may miss.

\finding{
{\bf Insights from RQ1: \ding{224}} Traditional feature-based models (TF-IDF, BoW) achieve reasonable recall but suffer from imbalanced precision, while transformer-based models like LongFormer and CodeBERT provide more robust detection by capturing long-range and contextual dependencies in IaC code. Fine-tuning and careful hyperparameter tuning further boost their effectiveness, making NLP-integrated models the preferable choice for detecting security misconfigurations in real-world IaC scripts.
}

\subsection{RQ2: Role of Semantic Analysis in Misconfiguration Detection}
\label{subsec:rq2}
To assess the importance of semantic context in ML-based static analysis of IaC scripts, we conducted a comparative evaluation using LongFormer and CodeBERT on our initial Ansible and Puppet datasets with full-context code snippets and on our ablated Puppet and Ansible datasets. Our goal was to examine how reduced contextual information, such as code and text (both forms of natural language)~\cite{10.5555/2337223.2337322}, influences the fine-tuned models' ability to identify security misconfigurations in IaC scripts.

We report in the following lines, the results from our ablation experiment derived from the evaluation of both fine-tuned models with full data context versus reduced data context. 

\subsubsection{LongFormer on Puppet: Full vs. Reduced Context}

We observe in Table~\ref{tab:longformer-ablation} that LongFormer performs better with full code context. On long Puppet snippets, the model achieves strong precision (0.87) and reasonable recall (0.75). When context is reduced, recall increases substantially to 0.97, but precision drops sharply to 0.55, leading to a lower F1-score (0.70 vs. 0.79).

\begin{table}[htbp]
\centering
\caption{Performance Comparison of Fine-Tuned LongFormer with Puppet Complete and Ablated Datasets for the Detection of Security Misconfigurations}
\label{tab:longformer-ablation}
\resizebox{\linewidth}{!}{
\begin{tabular}{lcc}
\hline
\textbf{Metric} & \textbf{Full Context} & \textbf{Reduced Context} \\ \hline
Precision & 0.87 & 0.55 \\
Recall    & 0.75 & 0.97 \\
F1-score  & 0.79 & 0.70 \\ \hline
\end{tabular}
}
\end{table}

This indicates that with reduced context, LongFormer becomes highly sensitive, detecting nearly all misconfigurations (high recall), but at the cost of many false positives (low precision), often flagging correct code as insecure. Without sufficient surrounding information, the model struggles to generalize and is more prone to over-reporting. These results highlight the importance of rich contextual information for LongFormer to accurately distinguish secure from misconfigured Puppet code.

\subsubsection{CodeBERT on Ansible: Full vs. Reduced Context}

We observe in Table~\ref{tab:codebert-ablation} that CodeBERT (batch size 128) performs substantially better when provided with full code context. With complete input, the model achieves a precision of 0.92 and recall of 0.88. In contrast, reducing context leads to a sharp decline in precision (0.46) and a moderate drop in recall (0.79), resulting in a much lower F1-score (0.58 vs. 0.90).

\begin{table}[htbp]
\centering
\caption{Performance Comparison of Fine-Tuned CodeBERT on Complete and Ablated Ansible Datasets for the Detection of Security Misconfigurations}
\label{tab:codebert-ablation}
\resizebox{\linewidth}{!}{
\begin{tabular}{lcc}
\hline
\textbf{Metric} & \textbf{Code Without Text} & \textbf{Code + Text} \\ \hline
Precision & 0.46 & 0.92 \\
Recall    & 0.79 & 0.88 \\
F1-score  & 0.58 & 0.90 \\ \hline
\end{tabular}
}
\end{table}

These results indicate that, although CodeBERT can still detect some misconfigurations under reduced context (moderate recall), it produces a high number of false positives (low precision), often mislabeling correct configurations as insecure. In other words, without sufficient semantic context, CodeBERT struggles to generalize effectively and tends to over-flag potential issues. This highlights the critical role of contextual information in enabling CodeBERT to accurately distinguish between secure and misconfigured IaC snippets.

\finding{
{\bf Insights from RQ2: \ding{224}} This experiment demonstrates that semantic context is not just helpful but essential for IaC misconfiguration detection in ML-based solutions: without it, models over-flag and lose precision, undermining practical usability. Future work could benefit from the combination of long-sequence modeling with semantic representations to build trustworthy security analysis tools.
}

\subsection{RQ3: LLMs Performance Comparison with CoderBERT and LongFormer}
\label{subsec:rq3}

We additionally conducted a comparative experiment between our approach and widely used large language models, evaluated on both the Ansible and Puppet datasets.
The results presented below include general-purpose LLMs (ChatGPT-4-Turbo, ChatGPT-3.5-Turbo, LLaMA3, and StarCoder2), which are trained on diverse text and code corpora, as well as code-specific models (CodeBERT and LongFormer) that were trained and fine-tuned specifically on IaC code.

On the Ansible dataset, as shown in Table~\ref{tab:llm_ansible_performance}, CodeBERT achieves the best overall performance with an F$_1$ score of \num{0.90}, combining high precision (\num{0.92}) and recall (\num{0.88}).

\begin{table}[htbp]
\centering
\small
\caption{Overall Performance of LLMs and LongFormer on Puppet Dataset for Security Misconfiguration Detection}
\label{tab:llm_puppet_performance}
\resizebox{\linewidth}{!}{
\begin{tabular}{lcccc}
\hline
\textbf{Model} & \textbf{Precision} & \textbf{Recall} & \textbf{F$_1$ Score} \\ \hline
ChatGPT-4 T    & 0.80  & 0.79  & 0.79  \\
ChatGPT-3.5 T  & 0.76 & 0.99 & 0.86 \\
LLaMA 3 8B     & 0.50 & 0.98 & 0.66 \\
StarCoder2 7B  & 0.68 & 1.00 & 0.81 \\
LongFormer  & 0.87 & 75 & 0.79 \\ \hline
\multicolumn{4}{l}{\small $^{\mathrm{*}}$ T = Turbo, B = Billion parameters.}
\end{tabular}
}
\end{table}

These findings suggest that code-specific pretraining remains advantageous for analyzing structured IaC configurations.
In contrast, GPT-3.5-Turbo attains an F1-score of \num{0.87}, with perfect precision (\num{1.00}) but lower recall (\num{0.77}).
While this indicates strong detection of true misconfigurations, it also highlights a risk of over-classification, where non-misconfigured scripts may be incorrectly flagged.
GPT-4-Turbo demonstrates a more balanced trade-off between precision (\num{0.79}) and recall (\num{0.73}), yet its F1-score (\num{0.73}) remains lower than both GPT-3.5-Turbo and CodeBERT.
Open-source LLMs, such as LLaMA~3~8B and StarCoder2~7B, exhibit extreme trade-offs: LLaMA~3 favors recall at the expense of precision, whereas StarCoder2 prioritizes precision but fails to capture many true positives.
This instability renders them less effective for reliable analysis of IaC scripts.

On the Puppet dataset, as shown in Table~\ref{tab:llm_puppet_performance}, our fine-tuned LongFormer demonstrates strong and consistent performance, achieving a precision of \num{0.87}, a recall of \num{0.75}, and an F1-score of \num{0.79}.
These results highlight the benefit of tailoring a transformer model to IaC scripts, effectively balancing detection accuracy while reducing false positives.
\begin{table}[htbp]
\centering
\small
\caption{Overall Performance of LLMs on Ansible Dataset for Security Misconfiguration Detection}
\label{tab:llm_ansible_performance}
\resizebox{.8\linewidth}{!}{
\begin{tabular}{lcccc}
\hline
\textbf{Model} &  \textbf{Precision} & \textbf{Recall} & \textbf{F1-score} \\ \hline
ChatGPT-4 T   & 0.79 & 0.73 & 0.73 \\
ChatGPT-3.5 T  & 1.00 & 0.77 & 0.87 \\
LLaMA 3 8B      & 0.48 & 1.00 & 0.64 \\
StarCoder2 7B  & 0.99 & 0.53 & 0.69 \\ 
CodeBERT  & 0.92 & 0.88 & 0.90 \\ \hline
\multicolumn{4}{l}{\small $^{\mathrm{*}}$ T = Turbo, B = Billion parameters.}
\end{tabular}
}
\end{table}

GPT-3.5-Turbo again prioritizes recall (\num{0.99}) at the expense of precision (\num{0.76}), resulting in an F1-score of \num{0.8662}.
GPT-4-Turbo exhibits a more balanced performance with precision of \num{0.80}, recall of \num{0.79}, and an F1-score of \num{0.79}, aligning more closely with LongFormer yet with slightly lower precision.
Open-source LLMs follow the same trend observed in the Ansible dataset, showing skewed performance profiles that constrain their practical deployment.

Together, these results illustrate a clear trade-off between the two model families:
\begin{itemize}[noitemsep,topsep=0pt,left=0pt]
    \item Code-specific transformers (CodeBERT, LongFormer) deliver balanced and interpretable performance, particularly excelling on Ansible where domain specialization enables effective detection of structured misconfigurations.
    \item General-purpose LLMs (GPT-3.5, GPT-4) provide broader detection coverage and exceptionally high recall (especially on Puppet), but often at the cost of reduced precision and elevated false positive rates.
    \item Open-source LLMs exhibit instability, with one-sided performance behaviors that render them less suitable without further fine-tuning.
\end{itemize}

\finding{
{\bf Insights from RQ3: \ding{224}} Our evaluation shows that general-purpose LLMs (e.g., GPT-3.5, GPT-4) achieve very high recall in detecting IaC security misconfigurations, but often at the cost of precision and increased false positives. In contrast, task-specific transformers like CodeBERT and LongFormer provide more balanced and stable performance, with CodeBERT excelling on Ansible and LongFormer performing strongly on Puppet. These results suggest that LLMs are powerful for broad detection coverage, while code-specific models remain essential for precision and reliability, highlighting the potential of hybrid approaches that combine both strengths.
}

%% file: threats.tex
\section{Discussion}
\label{sec:discussion}
We describe in this section the potential usefulness of AI-driven methods for the enhancement of software security, beyond the scope of IaC scripts, and further discuss the limitations and threats to validity inherent to our work.
\subsection{Potential of ML Models and LLMs for Software Security}
ML models and LLMs offer a transformative potential for enhancing software security. Traditional static analysis tools and rule-based systems, while valuable, often struggle to generalize beyond predefined patterns and may fail to detect novel or context-dependent vulnerabilities. In contrast, ML approaches can learn from large-scale, heterogeneous datasets, enabling them to recognize subtle correlations and patterns that are difficult to encode manually. LLMs, trained on vast corpora of code and natural language, further extend this capability by embedding both syntactic structures and semantic intent, allowing them to reason about code behavior in a more human-like manner. Their ability to process long sequences, capture cross-file dependencies, and adapt through fine-tuning makes them well-suited for detecting complex security flaws in IaC and other software artifacts. As software systems grow in scale, complexity, and interconnectivity, ML and LLM-based methods provide an adaptable, future-proof approach for preemptively identifying vulnerabilities and mitigating risks.

\subsection{Threats to Validity}
To ensure the reliability and reproducibility of our findings, we examine potential threats to validity across internal, external, and construct dimensions, and discuss the measures taken to mitigate them. 

\subsubsection{Threats to Validity on our Experimental Setup}
First, there may be internal validity concerns about whether the observed performance gains are genuinely due to our proposed approach rather than uncontrolled experimental factors. Threats may arise from dataset preprocessing, model fine-tuning instabilities, or hyperparameter settings that inadvertently favor CodeBERT or Longformer. To mitigate these risks, we applied a uniform preprocessing pipeline across all baselines, performed stratified data splits to preserve class balance, and used the same hyperparameter search methodology for all models. We avoided test-set leakage by tuning exclusively on fine-tuning/validation data and averaged results over multiple random seeds from our test dataset to reduce stochastic variance. Second, external validity concerns may relate to whether the findings generalize beyond our chosen datasets and IaC script types. Our evaluation focused on Ansible and Puppet scripts, so performance on other IaC languages (e.g., Terraform, Chef) remains to be tested. To mitigate this, we curated datasets containing diverse syntactic and semantic misconfiguration types and sourced scripts from multiple real-world repositories. The proposed approach is model-agnostic and transferable. Finetuning CodeBERT or Longformer on other types of IaC scripts should require minimal adaptation, supporting applicability to broader infrastructure contexts. Finally, the construct validity threats could relate to whether our experiments truly measure accurate detection of IaC security misconfigurations. Risks include noisy or incomplete annotations or narrow definitions of misconfigurations. To address these risks, we cross-checked labels against prior research datasets and validated all of them manually. We employed standard evaluation metrics, namely, precision, recall, and F1-score, to capture both correctness and completeness, and compared CodeBERT and Longformer directly against established static analysis and conventional ML baselines to ensure that observed improvements reflect genuine advances rather than metric artifacts.

\subsubsection{Threats to Validity on the use of LLMs}
While some recent works have raised concerns about prompting being potentially harmful or unreliable~\cite{morris2024prompting,perez2022ignore}, we emphasize that our use of system prompts is deliberate, controlled, and task-specific. In our study, system prompts are used primarily to perform small, deterministic automation tasks, such as removing comments from code, reducing code snippet length, or standardizing formatting. These tasks do not involve creative or unsafe content generation, and the prompts are designed to minimize ambiguity, ensuring reproducible results. Moreover, we combine prompting with post-processing manual validation to reduce the risk of unintended behavior. Hence, prompting is used as a safe and effective tool to assist static analysis and code normalization, rather than an open-ended generative activity. Furthermore, we evaluated LLMs for static analysis of IaC scripts (Ansible and Puppet scripts). Although LLMs capture code context, they do not outperform our fine-tuned CodeBERT and LongFormer models. One potential concern is data leakage from pretraining. However, our evaluation dataset is highly domain-specific, making overlap with LLM pretraining less probable. LLMs are also not explicitly optimized for detecting fine-grained security misconfigurations~\cite{li2025everything,yin2024multitask}, which reduces, but does not eliminate, the risk of memorization. Therefore, we put efforts into building a detection consensus for our selected LLMs on CWEs, whereas our fine-tuned models are fine-tuned on the target datasets. 

%% file: related.tex
\section{Related Work}
\label{sec:related}

Early research on IaC scripts' misconfigurations has largely focused on statistical code characteristics and process metrics. Rahman~\cite{rahman2018characterizing} identified structural and churn-related features, such as frequency of change and ownership, that correlate with IaC defects and violations of security principles like confidentiality and integrity. Subsequent work proposed Machine Learning frameworks for IaC misconfiguration prediction using product and process metrics across 104 repositories~\cite{dalla2021within}, finding that product-based features better identify misconfigurations, and that models such as Random Forest perform well in classification tasks. However, these approaches treat IaC scripts as flat collections of tokens, lacking deeper semantic understanding. In contrast, other works have started to leverage natural language with ML~\cite{borovits2022findici} to show inconsistencies in IaC scripts using code descriptions (task names, code comments) and code instructions. In this work, we consider both code and text as natural language as defined in the literature~\cite{10.5555/2337223.2337322}. Our work differentiates itself from these studies by considerong both code and text as forms of natural language. Furthermore, we leverage CodeBERT for syntax-aware encoding and Longformer for handling the extended length of long IaC scripts, to enrich the representation of IaC code descriptions and IaC code-only for the detection of security misconfigurations. By doing so, we bridge more syntactic and semantic cues for security misconfiguration detection, addressing the limitations of prior statistical and ML-based methods for the analysis of IaC scripts.

%% file: conclusion.tex
\section{Conclusion}
\label{sec:conclusion}

Through extensive experimentation on real-world datasets of misconfigured IaC scripts, this study proposes an approach that leverages semantic code understanding and richer context representation of long IaC code sequences. Specifically, we propose a novel approach that treats IaC scripts as natural language, rather than limiting the analysis to their programming-language aspects, as is common in prior work. We implement our work with fine-tuned ML models such as CodeBERT and Longformer, and it substantially outperforms traditional static analysis and conventional ML-based techniques in detecting IaC security misconfigurations. The results highlight the effectiveness of combining NLP and code-aware modeling to capture nuanced intent and contextual dependencies in IaC scripts. Beyond the immediate performance gains, this work hints at LLM-powered analysis as a flexible, adaptable, and future-proof framework for securing evolving infrastructure environments, uniting the strengths of NLP and code processing.